\documentclass[11pt]{article}
\usepackage{graphicx}

\title{Parallel Supercomputing PC Cluster and 
Some Physical Results in Lattice QCD\thanks{This work 
is supported by the
National Science Fund for Distinguished Young Scholars (19825117),
Key Project of National Science Foundation (10235040), 
Guangdong Provincial Natural Science Foundation, 
National and Guangdong Ministries of Education, 
Ministry of Science and Technology,
Foundation of Zhongshan University Advanced Research Center,
and Guoxun (Guangdong National Communication Network) Ltd.}}

\vspace{2cm}
\author{LUO Xiang-Qian$^1$, MEI Zhong-Hao$^1$,  Eric B. GREGORY$^1$,  \\ 
YANG Jie-Chao$^2$, WANG Yu-Li$^2$, and LIN Yin$^2$\\
$^1${\small\sl Department of Physics, Zhongshan University, 
Guangzhou 510275, China}\\
$^2${\small\sl Guangdong National Communication Network Science and Technology Stock Co., Ltd,}\\ 
{\small\sl Guangzhou 510080, China} \\}

\date{\today}

\begin{document}
\maketitle

\begin{abstract}
We describe the construction of 
a high performance parallel computer composed of PC components,  
present some physical results for light hadron and hybrid meson masses from lattice QCD. We also show that the smearing technique is very useful for improving the spectrum calculations.
\end{abstract}
\centerline{{\bf Keywords}: Computational Physics, Parallel Computing, Lattice QCD}

\setcounter{page}{0}
\newpage

\section{Introduction}

The interests of the computational physics and high energy physics group\cite{zsusite} at the Zhongshan University (ZSU) cover such topics
as lattice gauge theory\cite{Luo:1996tx,Luo:1996ha,Luo:1996sa,luo,Luo:2001id, Gregory:1999pm,Luo:2000xi,Huang:1999fn}, supersymmetry\cite{Catterall:2000rv}, 
quantum instantons\cite{Jirari:1999bx} and quantum chaos\cite{Jirari:1999ij, Caron:2001zb}. 
All of these topics can be 
investigated through Monte Carlo simulation, but can be quite costly in 
terms of computing power.  In order to do large scale 
numerical investigations of these topics, we require a corresponding 
development of our local computing resources.

The last two decades have ushered in the computer revolution for 
the consumer.  In this period computers have moved from the domain of 
large companies, universities, and governments, to private homes and small 
businesses.  As computational power has become more accessible, our demands 
and expectations for this power have increased accordingly.  

We demand an 
ever-increasing amount of computational ability for business, communication, 
entertainment, and scientific research.  This rapid rise in both the demand 
for computational ability as well as the increase of that capability itself
have forced a continual redefinition of the concept of a ``super computer.''
The computational speed and ability of household computers now surpasses 
that of computers which helped guide men to the moon.  The demarcation 
between super computers and personal computers has been 
further blurred in recent years by the high speed and low price of modern 
CPUs and networking technology and the availability of low cost or free
software.  By combining these three elements - all readily 
available to the consumer - one can assemble a true super computer that is
within the budget of small research labs and businesses.  This type of 
cluster is generally termed a Beowulf class computer. The idea was originally
developed as a project at the US National Aeronautics and Space 
Administration\cite{beowulf}.

  We document the construction of a cluster of PCs, configured 
to be capable of parallel processing, and show the performance in lattice QCD simulations.  We also present some results for the hadron masses from lattice QCD.

\begin{figure}[hb]
\begin{center}
\rotatebox{270}{\includegraphics[width=8cm]{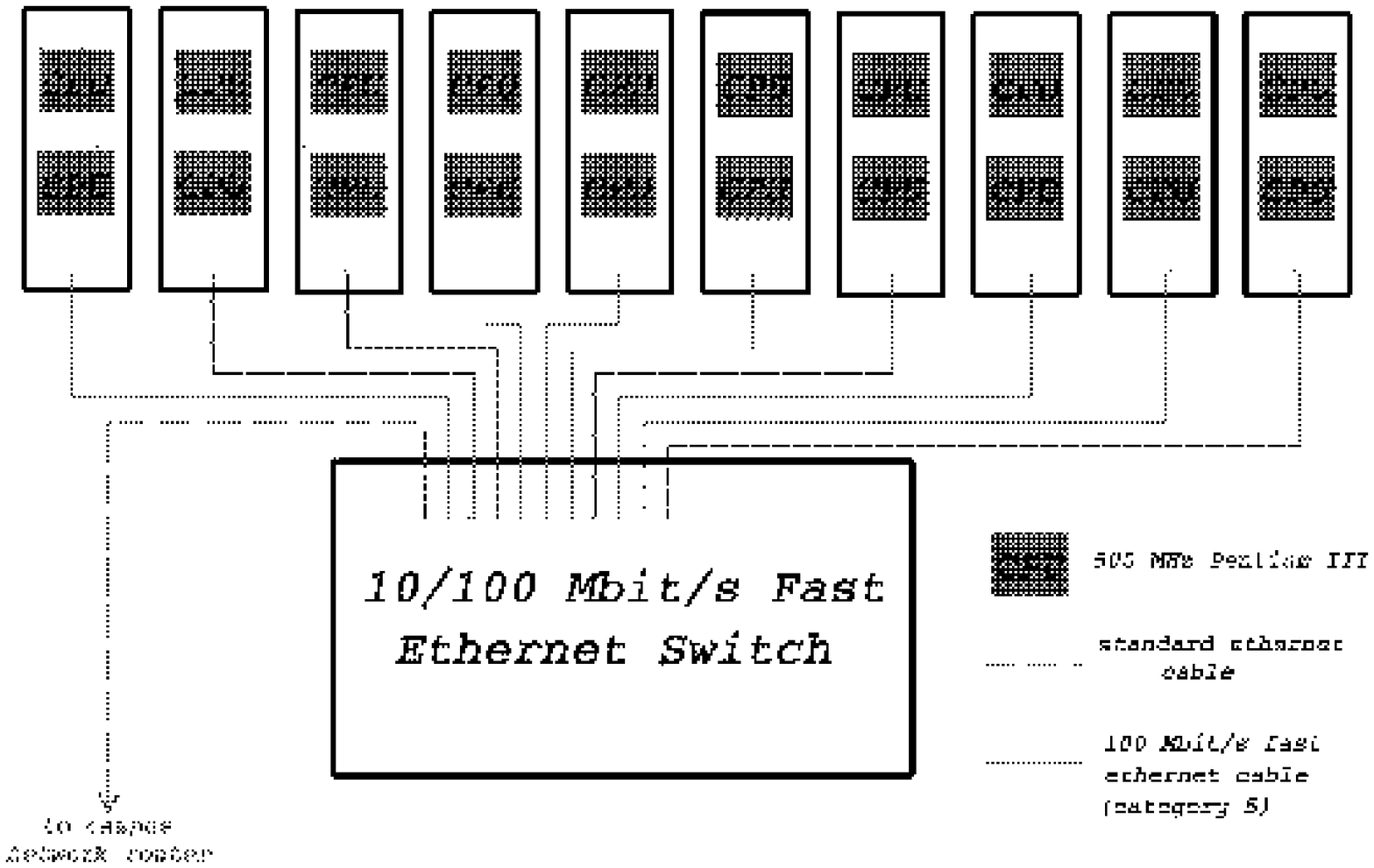}}
\end{center}
\caption{\label{diagram} Schematic diagram of a parallel cluster.}
\end{figure}

\section{Construction of a Parallel Cluster}

\subsection{Computational Hardware}

We built a cluster of 10 PC type computers, all the components of which 
we purchased at normal consumer outlets for computer equipment.  The major 
difference in our computers from one likely to be found in a home or business
is that each is equipped with two CPUs.  This allows us to roughly double 
our processing power without the extra overhead cost for extra cases, power 
supplies, network cards, etc.  Specifically, we have installed two 500MHz
Pentium III processors in each motherboard.  For the purposes of this report 
we will describe each computer as one ``node'' in the cluster; i.e., a node has
two processors.  Each node has its own local EIDE hard disk, in our case each 
has 10GB.  This amount of space is not necessary,
as the operating system requires less than one gigabyte per node, however
the price of IDE hard disks has dropped so rapidly that it seems a reasonable 
way to add supplementary storage space to the cluster.  
Furthermore, each node is equipped with memory (at least 128MB), a display 
card, a 100Mbit/s capable fast Ethernet card, a CDROM drive and a floppy 
drive.  
These last two items are not an absolute necessity as installation can be 
done over the network, but they add a great deal of convenience and 
versatility for a very modest cost.

One node is special and equipped with extra or enhanced components.  The 
first node acts as a file server and has a larger (20GB) hard disk.
This disk is the location of all the home directories associated with user 
accounts.  The first node also has a SCSI adapter, for connecting external 
backup devices such as a tape drive.  

What each computer does not have is a monitor, keyboard, and mouse.  Monitors 
can easily be one of the most expensive components of a home computer system.
For a cluster such as this one, the individual nodes are not intended for 
use as separate workstations. Most users access the cluster through
network connections.  We use a single console (one small monitor, 
a keyboard and mouse) for administrative tasks. It is handy when installing 
the operating 
system on a new node.  In this situation we move the console cables to the 
particular node requiring configuration. Once we have installed communications
programs such as telnet and ssh, it is almost never necessary to move the 
monitor and cables to the subordinate nodes.

\subsection{Communications Hardware}

There are many options for networking a cluster of computers, including 
various types of switches and hubs, cables of different types and 
communication protocol.  We chose to use fast Ethernet technology, as a 
compromise between budget and performance demands.  We have already stated 
that we equipped each node with a 100Mbit/s capable fast Ethernet card. 
A standard Ethernet hub has the limitation on not being able to accommodate
simultaneous communications between two separate pairs of computers, so we use
a fast Ethernet switch.  This is significantly more expensive than a hub, but 
necessary for parallel computation involving large amounts of inter-node 
communication.  We found a good choice to be a Cisco Systems 2900 series 
switch.  For ten nodes a bare minimum is a 12 port switch: one port for each 
node plus two spare ports for connecting either workstations or a connection 
to an external network.  We have in fact opted for a 24 port switch to leave
room for future expansion of the cluster as our budget permits.

100Mbit per second communication requires higher quality ``Category-5'' 
Ethernet 
cable,  so we use this as the connection between the nodes and the switch.
It should be noted that while a connection can be made from one of the switch
ports to an external Internet router, this cable must be ``crossover'' cable
with the input and output wire strands switched.  
The general layout of the cluster hardware is shown in Figure \ref{diagram}.

\subsection{Software}

For our cluster we use the Linux open source UNIX-like operating system. 
Specifically, we have installed a Redhat Linux distribution, due to the 
ease of installation.  The most recent Linux kernel versions automatically 
support dual CPU computers.  Linux is also able to support a 
Network File System (NFS), allowing all of the nodes in the cluster to share
hard disks, and a Network Information System (NIS), which standardizes the
usernames and passwords across the cluster.  

The one precaution one must take before constructing 
such a cluster is that the hardware components are compatible with Linux.
The vast majority of PC type personal computers in the world are running a 
Windows operating system, and hardware manufacturers usually write only 
Windows device drivers.  Drivers for Linux are usually in the form of 
kernel modules and are written by Linux developers.  As this is a distributed 
effort, shared by thousands of programmers worldwide, often working as 
volunteers, every PC hardware component available 
is not necessarily immediately compatible with Linux. Some 
distributions, 
such as Redhat have the ability to probe the hardware specifications  
during the installation procedure.  It is rather important to check on-line 
lists of compatible hardware --- particularly graphics cards and network cards 
--- before purchasing hardware.  We began by purchasing one node first and 
checking the compatibility with the operating system first before purchasing 
the rest of the nodes.

To provide parallel computing capability, we use an Message Passing Interface
(MPI) implementation.  MPI is a standard specification for message passing 
libraries\cite{mpi}.  Specifically we use the {\em mpich} implementation, which
is available for free download over the world wide web\cite{mpich}.
An MPI implementation is a collection of software that allows communication 
between programs running on separate computers.  It includes a library of 
supplemental
{\em C} and {\em FORTRAN} functions to facilitate passing data between the 
different processors.

\section{Basic Ideas of Lattice QCD}

Our main purpose for building the PC cluster is to do large scale 
lattice Quantum Chromodynamics (QCD) simulations.
The basic idea of lattice gauge theory\cite{Wilson:1974sk}, 
as proposed by K. Wilson in 1974, is to replace the continuous 
space and time by a discrete grid:

\begin{center}
\setlength{\unitlength}{.02in}
\begin{picture}(80,80)(0,0)

\multiput(25,5)(10,0){6}{\circle*{2}}
\multiput(25,15)(10,0){6}{\circle*{2}}
\multiput(25,25)(10,0){6}{\circle*{2}}
\multiput(25,35)(10,0){6}{\circle*{2}}
\multiput(25,45)(10,0){6}{\circle*{2}}
\multiput(25,55)(10,0){6}{\circle*{2}}

\put(11,30){\vector(0,2){25}}
\put(11,30){\vector(0,-2){25}}
\put(8,30){\makebox(0,0){$L$}}

\put(50,55){\vector(1,0){5}}
\put(50,55){\vector(-1,0){5}}
\put(50,60){\makebox(0,0)[t]{$a$}}

\put(25,63){\vector(0,-1){8}}
\put(30,70){\makebox(0,0)[r]{site}}

\put(75,45){\vector(0,1){10}}
\put(90,50){\makebox(0,0)[r]{link}}
\put(75,45){\line(0,1){10}}

\put(65,5){\vector(1,0){10}}
\put(75,5){\vector(0,1){10}}
\put(75,15){\vector(-1,0){10}}
\put(65,15){\vector(0,-1){10}}
\put(115,10){\makebox(0,0)[r]{plaquette}}
\end{picture}
\end{center}

\vskip 0.5cm

\noindent
Gluons live on links $U(x,\mu)=e^{-ig\int_x^{x+ \hat{\mu}a} dx' A_{\mu}(x')}$, and quarks live on lattice sites.  
The continuum Yang-Mills action $S_g=\int d^4x ~\rm{Tr} F_{\mu \nu}(x)F_{\mu \nu}(x)/2$ is replaced by
\begin{eqnarray}
S_g=-{\beta \over 6} \sum_p \rm{Tr} (U_{p} +U_{p}^{\dagger}-2),
\label{gauge}
\end{eqnarray}
where $\beta=6/g^2$, and  $U_p$ is the ordered product of link variables $U$ around an elementary 
plaquette.  The continuum quark action 
$S_q=\int d^4x ~\bar{\psi^{cont}} (x) (\gamma_{\mu}D_{\mu}+m) \psi^{cont}(x)$  is replaced by
\begin{eqnarray}
S_q &=& \sum_{x,y} \bar{\psi}(x) M_{x,y}\psi(y).
\label{quark}
\end{eqnarray}
For Wilson fermions, the quark field $\psi$ on the lattice is related to the continuum one $\psi^{cont}$ 
by $\psi=\psi^{cont} a^3/(2\kappa)$ with $\kappa=1/(2ma+8)$. $M$ is the fermionic matrix:
\begin{eqnarray}
    M_{x,y}=\delta_{x,y}-
\kappa \sum_{\mu=1}^{4} \left[ (1-\gamma_{\mu})U_{\mu}(x)\delta_{x,y-\hat{\mu}}
+
(1+\gamma_{\mu})U_{\mu}^{\dagger}(x-\hat{\mu})\delta_{x,y+\hat{\mu}} \right].
\end{eqnarray}
For Kogut-Susskind fermions, the fermionic matrix is given by
\begin{eqnarray}
M_{x,y}  &=& ma \delta_{x,y}             
+
{1 \over 2}\sum_{\mu=1}^{4} \eta_{\mu}(x)  \left[ U_{\mu}(x) \delta_{x,y-\hat{\mu}}
-U_{\mu}^{\dagger} (x-\hat{\mu}) \delta_{x,y+\hat{\mu}} \right],
\nonumber \\
\eta_{\mu}(x) &=& (-1)^{x_1+x_2+...+x_{\mu-1}}.
\label{KS}
\end{eqnarray}

Physical quantities are calculable through 
Monte Carlo (MC) simulations with importance sampling.
Fermion fields must be integrated out before the simulations, leading to
\begin{eqnarray}
\langle F \rangle = {\int [dU] \bar{F}([U]) e^{-S_{g}([U])} 
\left( \det M \right)^{N_f} 
\over \int [dU] e^{-S_g([U])}\left( \det M \right)^{N_f}}
\approx {1 \over N_{\rm config}} \sum_{\mathcal Conf}  \bar{F}[{\mathcal Conf}]. 
\end{eqnarray}
Here $\bar{F}$ is the operator after Wick contraction of the fermion fields
and the summation is over the gluonic configurations, ${\mathcal Conf}$, drawn 
from the Boltzmann distribution.  In quenched approximation, $\det (M)=1$.

We introduce the $u$ and $d$ quark propagators
\begin{eqnarray}
   Q^u_{s_1c_1,s_2c_2}(x,y)&=&M^{-1}[U,\kappa=\kappa_u]_{{xs_1c_1},ys_2c_2},
\nonumber \\
   Q^d_{s_1c_1,s_2c_2}(x,y)&=&M^{-1}[U,\kappa=\kappa_d]_{{xs_1c_1},ys_2c_2},
\label{prop}
\end{eqnarray}
where the Dirac and color indexes are explicitly written.
In general, most 
of the computer time in the simulation of hadron masses or dynamical quarks 
is spent
on the computations of the quark propagators.
Usually these operations are accomplished by means of
some inversion algorithm, which solves linear equation
systems.

To compare with the real world, 
the continuum limit $a \to 0$ should be eventually taken. 
On the other hand, to keep the physical volume $(La)^3$ unchanged, 
the number of spatial lattice sites $L^3$ should be very large.
To reliably measure the effective mass of a hadron,
one has also to increase the number of temporal lattice sites $T$ accordingly. 
Therefore, the computational task will then be tremendously increased. 
As such, it is well suited for parallelization.  A parallel 
lattice QCD algorithm divides the lattice into sections and assigns the 
calculations relevant to each section to a different processor.  Near the 
boundaries of the lattice sections, information must be exchanged between 
processors. However, since the calculations are generally quite local, the 
inter-processor communication is not extremely large.

\begin{figure}[hb]
\begin{center}
\includegraphics[width=10cm]{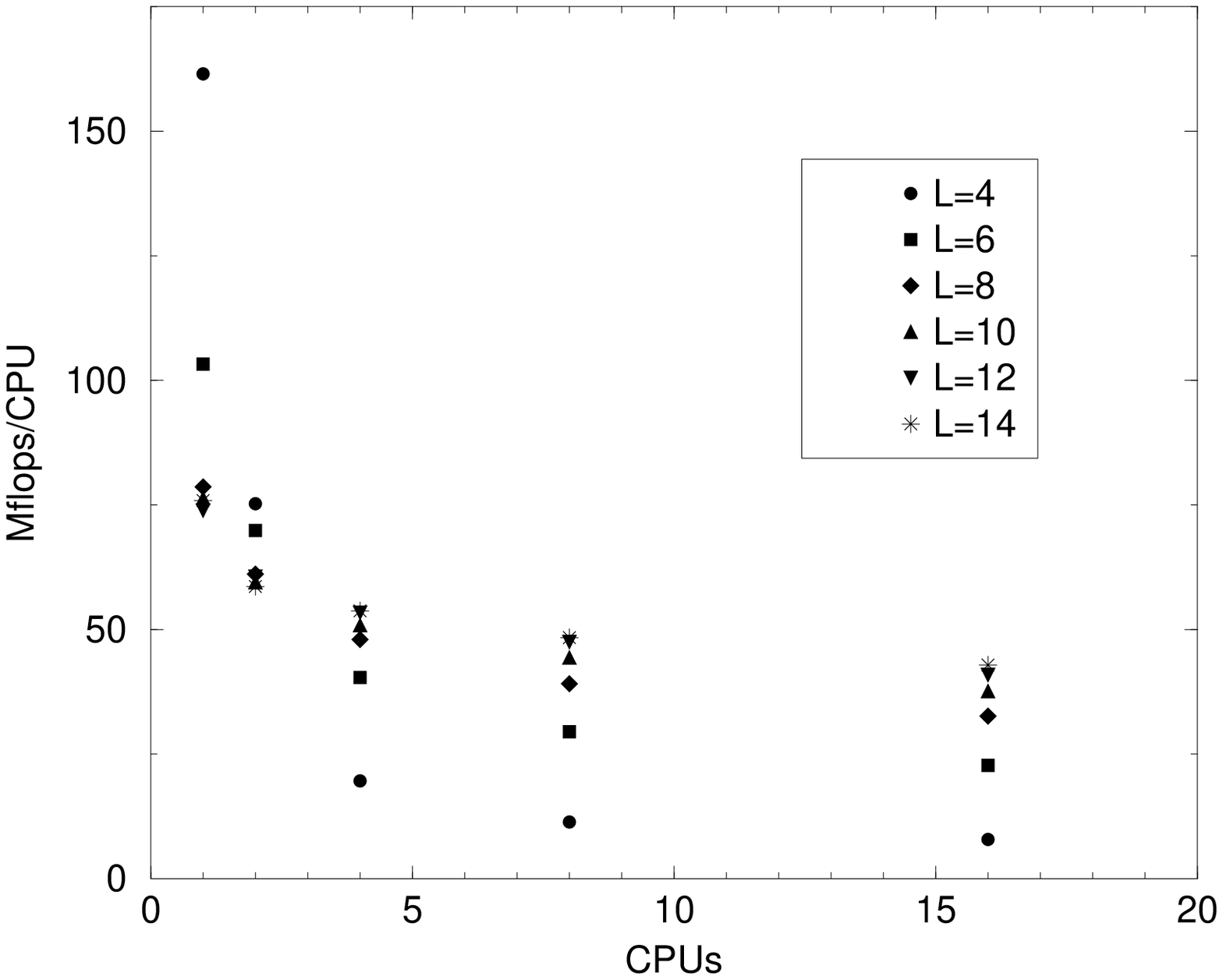}
\end{center}
\caption{\label{milc_perf} Performance (in Mega-Flops) per CPU versus the 
number of CPUs in the MILC QCD code benchmark.}
\end{figure} 	 

\section{Performance and Cost}

We ran a standard LINPACK benchmark test and determined the peak speed of a 
single 500MHz Pentium III processor. The results of this test are shown in 
Table \ref{tab1} to be about 100 million floating point 
operations a second (Mflops). With this in mind, we can say that the 
theoretical upper 
limit for the aggregate speed of the whole cluster (20 CPUs) approaches 
2 Gflops.  Of course this is possible only in a computational task 
that is extremely parallelizable with minimum inter-node communications, no
cache misses, etc.  In the year 2000, the cost for our entire cluster was 
about US\$15,000, including the switch.  
This means that the cost for 
computational speed was about 	US\$7.50/Mflop.  (Eliminating less essential 
hardware such are CDROMS, display cards, and floppy drives and using smaller 
hard disks on the subordinate nodes would further reduce this number.)   
It is instructive to compare 
this to other high performance computers. One example is a Cray T3E-1200.
Starting at  US\$630,000 for six 1200 Mflop processors\cite{craysite}, the 
cost is about US\$87.50 per Mflop.  The Cray is more expensive by an order of 
magnitude. Clearly there are advantages in communication speed and other 
performance factors in the Cray that may make it more suitable for some types 
of problems.  However, this simple calculation shows that PC clusters are 
an affordable way for smaller research groups or commercial interests to 
obtain a high performance computer.

\begin{table}
\caption{Results of LINPACK benchmark test on a single CPU. \label{tab1}}
\begin{center}
\begin{tabular}{|c|c|}
\hline
{\bf Precision} &  {\bf Mflop} \\
\hline
single  & 86 - 114 \\
double  & 62 - 68 \\
\hline
\end{tabular}
\end{center}
\end{table}

\begin{figure}[hb]
\begin{center}
\includegraphics[width=11cm]{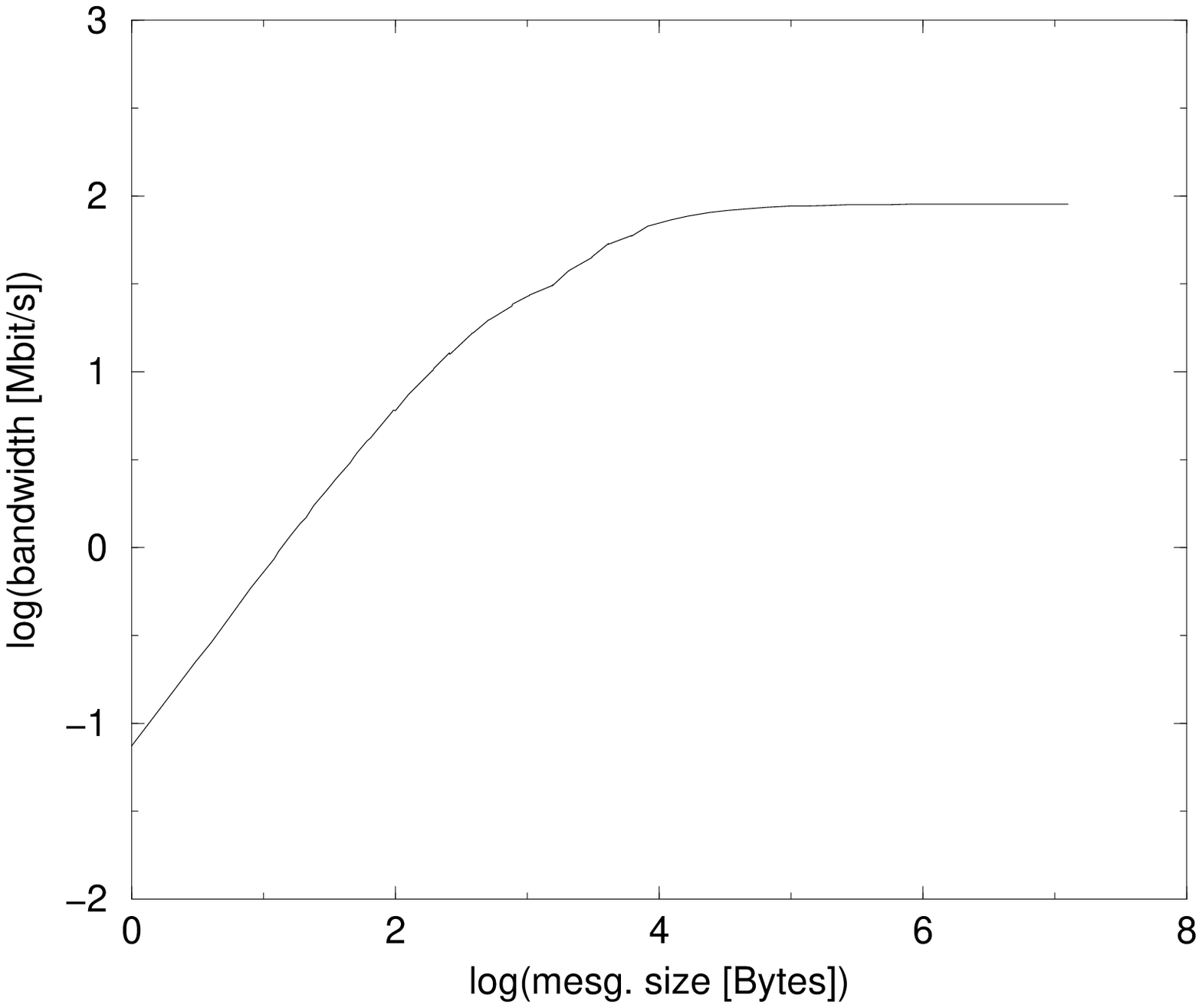}
\end{center}
\caption{\label{bandwidth} Communications bandwidth vs. message size.}
\end{figure} 	 

A widely used lattice QCD simulation program is the MILC (MIMD Lattice 
Collaboration) code \cite{milc}. It has timing routines provided so that 
one can use the parallelized conjugate gradient (CG) routine for
inverting the fermionic matrix in the simulation 
as a benchmark.
Furthermore, as this code is very versatile and is designed to be run on a 
wide variety of computers and architectures.  This enables quantitative 
comparison of our cluster to both other clusters and commercial supercomputers.
In the MILC benchmark test we ran to a convergence tolerance of $10^{-5}$ per
site.  For consistency with benchmarks performed by others, we simulated 
Kogut-Susskind fermions given by Eq. (\ref{KS}).

\begin{table}
\begin{center}
\begin{tabular}{|c|c|c|c|c|}
\hline
Interface directions & hypercubes (CPUs) & Lattice volume & 
Total interface  & interface / CPU\\
\hline
$j$  &  $2^j$  &  $L^{4-j} \times (2L)^j$  & $2^jjL^{3}$ & $jL^{3}$\\
\hline
0 & 1 & $L^4$ & 0 & 0 \\
1 & 2 & $L^3 \times 2L $ & $2L^3$ &  $L^3$\\
2 & 4 & $L^2 \times (2L)^2 $ & $8L^3$ & $2L^3$\\
3 & 8 & $L \times (2L)^3 $ & $24L^3$  & $3L^3$\\
4 & 16 & $(2L)^4 $ & $64L^3$  & $4L^3$\\
\hline
\end{tabular}
\end{center}
\caption{ \label{boundary} Summary of boundary sizes for division of a lattice
into 1, 2, 4, 8 and 16 hypercubes of size $L^4$.}
\end{table}

We illustrate the result of the MILC code benchmark test in 
Figure \ref{milc_perf}. This figure deserves some explanation. We have run the 
benchmark test for different size lattices and different numbers of processors.
It is useful to look at how performance is affected by the number of CPUs, 
when the amount of data per CPU is held fixed, that is each CPU is 
responsible for a section of the lattice that has $L^4$ sites. For one CPU,
the size of the total lattice is $L^4$. For two CPUs it is $L^3\times 2L$.
For four CPUs the total lattice is $L^2 \times (2L)^2$; for eight CPUs, 
$L \times (2L)^3$, and for 16 CPUs the total size of the lattice is $(2L)^4$.

Note that the falloff in performance with increased number of CPUs is dramatic.
This is 
because inter-processor message passing is the slowest portion of this or any
MPI program and from two to sixteen CPUs, the amount of communication per
processor increases by a factor of four. 
Table \ref{boundary} shows that for a lattice divided into $2^j$
hypercubes, each of size $L^4$, there will be $j$ directions in which the 
CPUs must pass data to their neighbors. The amount of communication each 
processor must perform is proportional to the amount of interface per 
processor. As this increases, per node performance decreases until $j=4$ and 
every lattice dimension has been divided (for a $d=4$ simulation), and the
per-processor performance should remain constant as more processors are added.
The shape of this decay is qualitatively consistent with $1/j$ falloff.

Of course there are other ways to divide a four-dimensional lattice. 
The goal of a particular simulation will dictate the geometry of the lattice 
and the therefore the most efficient way to divide it up (generally minimizing 
communication).  A four-CPU simulation using a $4L\times L^3$ lattice has 
the four hypercubic lattice sections lined up in a row (as opposed to in a 
$2\times 2$ square for a $L^2 \times (2L)^2$ lattice) and has the same amount 
of communication per CPU as does the $L^3\times 2L$
two-CPU simulation.  In a benchmark test the per-CPU performance was
comparable to the performance in the two-CPU test.

For a single processor, there is a general decrease in performance as $L$ 
increases, as shown in Tab. \ref{one_cpu}.
This is well explained in \cite{Gottlieb:2000dv} as due to the larger matrix size
using more space outside of the cache memory, causing slower access time to 
the data. 

\begin{table}
\begin{center}
\begin{tabular}{|c|c|}
\hline
$L$ & single processor speed (Mflops)\\
\hline
4 & 161.5\\
6 & 103.2\\
8 & 78.6\\
10 & 76.4\\
12 & 73.9\\
14 & 75.9\\
\hline
\end{tabular}
\end{center}
\caption{ \label{one_cpu} Summary of single CPU performance.}
\end{table}

For multiple CPUs there is in performance improvement as $L$ is increased.
The explanation for this is that the communication bandwidth is not constant 
with respect to message size, as Fig. \ref{bandwidth} shows. For very small 
message sizes, the bandwidth is very poor. It is only with messages of around 
10kB or greater that the bandwidth reaches the full potential of the fast 
Ethernet hardware, nearly 100Mbit/sec. With a larger $L$, the size of the
messages is also, improving the communication efficiency. The inter-node
communication latency for our system is 102$\mu s$. As inter-node 
communication is the slowest part, a parallel program this far out-ways the 
effect of cache misses.

\section{Physics Results}

\subsection{Green functions} 

Calculation of hadron spectroscopy remains to be an important task 
of non-perturbative studies of QCD using lattice methods. 
In this paper, 
we will present the spectrum results of light hadrons and $1^{-+}$ hybrid
meson with quenched Wilson fermions. 
$\pi$, $\rho$, proton or $\Delta^{++}$ consists of quark and anti-quark, 
and their operators are given by:
\begin{eqnarray}
 O^{\pi^+}(x) &=& \bar{d}_{s_1c}(x) \gamma_{5,s_1s_2}u_{s_2c}(x),
\nonumber \\ 
O_k^{\rho^+}(x) &=& \bar{d}_{s_1c}(x)\gamma_{k,s_1s_2}u_{s_2c}(x),
\nonumber \\
O_{s_1}^{p}(x) &=& \epsilon_{c_1 c_2 c_3}(C\gamma_{5})_{s_2 s_3} u_{s_1 c_1}(x)
\left( u_{s_2 c_2}(x)d_{s_3 c_3}(x)-d_{s_2 c_2}(x)u_{s_3 c_3}(x) \right),
\nonumber \\ 
O^{\Delta^{++}}_{s_1}(x) &=& 
\epsilon_{c_1c_2c_3}(C\gamma_{5})_{s_2s_3}u_{s_1c_1}(x)
u_{s_2c_2}(x)u_{s_3c_3}(x),
\label{operator}
\end{eqnarray}  
where $u$ and $d$ are the ``up'' and ``down'' quark fields, $C$ is the 
charge conjugation matrix, $c$ is the color index of the Dirac 
field, and $s$ is the Dirac spinor index. Summation over repeated index is implied. The correlation functions of a hadron is:
\begin{eqnarray}
   C_{h}(t)=\sum\limits_{\stackrel{\rightarrow}{x}}
\langle O_h^{\dagger}(\stackrel{\rightarrow}{x},t)O_h(0,0)\rangle,
\label{correlation_fun}
\end{eqnarray}
where $O_h(\stackrel{\rightarrow}{x},t)$  
is a hadron operator given in (\ref{operator}).
Then,
\begin{eqnarray}
   C_{\pi^+}(t)&=&-\langle \sum_{\vec{x}}{\rm Tr}_{sc} 
\left( \gamma_{5}Q_d(x,0)\gamma_{5}Q_u(0,x)\right)\rangle,
\nonumber \\
  C_{\rho^+}(t)&=&-\langle \sum_{\vec{x}}{\rm Tr}_{sc} 
\left( \gamma_{k}Q_d(x,0)\gamma_{k}Q_u(0,x)\right)\rangle,
\nonumber \\
  C_{p}(t)&=&\epsilon_{c_1c_2c_3}\epsilon_{c_4c_5c_6}
\left(C\gamma_{5}\right)_{s_3s_4}
\left(C\gamma_{5}\right)_{s_5s_6}
\left(
Q^u_{s_1c_1,s_2c_4}(x,y)
Q^u_{s_3c_2,s_5c_5}(x,y)
Q^d_{s_4c_3,s_6c_6}(x,y)\right.
\nonumber \\
&+&\left.
Q^u_{s_1c_1,s_5c_4}(x,y)
Q^u_{s_3c_2,s_2c_5}(x,y)
Q^d_{s_4c_3,s_6c_6}(x,y)\right),
\nonumber \\
C_{\Delta^{++}}(t)&=&
\epsilon_{c_1c_2c_3}\epsilon_{c_4c_5c_6}
\left(C\gamma_{k}\right)_{{s_3}{s_4}}
\left(C\gamma_{k}\right)_{{s_5}{s_6}}
\left(
Q^u_{{s_1c_1},s_2c_4}(x.y)
Q^u_{s_3c_2,s_5c_5}(x,y)
Q^u_{s_4c_3,s_6c_6}(x,y)\right.
\nonumber \\
&+&\left.
2Q^u_{s_1c_1,s_5c_4}(x,y)
Q^u_{s_3c_2,s_2c_5}(x,y)
Q^u_{s_4c_3,s_6c_6}(x,y)\right),
\nonumber \\
\label{correlation}
\end{eqnarray}
where $\rm{Tr}_{sc}$ stands for a trace over spin and color.
In  Tab. \ref{tab0}, we list the operator for the P-wave $a_1$ meson, 
which is also made of quark and anti-quark.

Hybrid (exotic) mesons, which are important predictions of quantum chromodynamics (QCD), 
are states of quarks and anti-quarks bound by excited gluons.
First principle lattice study of such states 
would help us understand the role of ``dynamical'' color in low energy QCD
and provide valuable information for experimental search for these new particles. 
In Tab. \ref{tab0}, the operator of $1^{-+}$ meson is given.

\begin{table} 
  \begin{center}
  \begin{tabular}{|c|c|c|c|}\hline
 meson         &  $J^{PC}$ &  Mnemonic      & Operator                    \\ \hline
    ${\bar q} q$     &           &                &                             \\
$f_1({\rm P-wave})$ & $1^{++}$  & $^3P_1 ~ f_1$  & $\epsilon_{ijk}{\bar \psi}\gamma_{j}\stackrel{\leftrightarrow}{\partial_{k}}\psi$    \\ \hline
   ${\bar q} q g$    &           &                &                               \\
   $1^{-+}$         & $1^{-+}$  & $\rho\bigotimes B$ & ${\bar \psi}^{c_1}\gamma_{j}\psi^{c_2}F^{c_1c_2}_{ji}$   \\
   $q^{4}$          &   $1^{-+}$ & $\pi \bigotimes a_{1}$  & ${\bar \psi}^{c_1}_{f_1}(\vec{x})\gamma_{5}
      \psi_{f_2}^{c_1}(\vec{x}){\bar \psi}^{c_2}_{f_2}(\vec{y})\gamma_{5}\gamma_{i}\psi^{c_2}_{f_3}(\vec{y})$  \\ 
\hline       
\end{tabular}
\end{center}
\caption{\label{tab0} Source and sink operators for $a_1$ and hybrid mesons.}
\end{table}

For sufficiently large values of $t$ and the lattice time period $T$,
the correlation function is expected to approach the asymptotic form:
\begin{eqnarray}
   C_h(t)\rightarrow Z_{h}[\exp(-m_{h}at)+\exp(m_{h}at-m_{h}aT)].
\label{fit}
\end{eqnarray} 
Fitting the equation at large $t$, the effective mass of a hadron $am_h$ is  
obtained.

\subsection{Light hadron masses}

We updated the pure SU(3) gauge fields
with Cabibbo-Marinari quasi-heat bath algorithm, 
each iteration followed by 4 over-relaxation sweeps.
The simulation parameters are listed in Tab. \ref{parameters}.
The distance between two nearest stored configurations is 100.
The auto-correlation time was computed 
to make sure that these configurations  are independent.

\begin{table}
\begin{center}
\begin{tabular}{|c|c|c|c|}\hline
  volume          &  $\beta$ & warmup & stored configs.  \\ 
\hline
   $8^3\times 32$ &    5.7   &    200  & 200       \\ 
\hline
   $8^3\times 32$ &    5.85  &    200  & 200        \\ 
\hline
  $12^3\times 36$ &    6.25  &    200  & 200         \\ 
\hline  
$16^3\times 32$   &    6.25  &    600  & 600         \\ 
\hline               
\end{tabular}
\end{center}
\caption{\label{parameters} Simulation parameters}
\end{table}

The $u$ quark are $d$ quark are assumed to be degenerate.
Using the CG algorithm, the quark propagators in Eq. (\ref{prop}) 
are calculated by
inverting the Dirac matrix with preconditioning via
ILU decomposition by checkerboards. The convergence tolerance we set 
is $5\times 10^{-8}$.
To extract masses from the hadron propagators, we must average the 
correlation function in Eq. (\ref{correlation}) 
of the hadron over the ensemble of gauge configurations, 
and use a fitting routine to evaluate $am_h$ in Eq. (\ref{fit}).

The quenched simulations were  performed at lattice coupling
of  $\beta=5.7$, $ \beta=5.85$  on the  $8^3\times32$ lattice. We compared 
the  results with those by MILC and GF11. 
At $\beta=6.25$,  we computed the light meson and baryon masses on the  
$12^3\times36$ and $16^3\times 32$ lattices. 
The data for $\beta=6.25$ have been reported in Ref. \cite{Mei:2002fu}.
Here we detail the results for $\beta=5.7$ and $\beta=5.85$.

\begin{figure}[hb]
\begin{center}
\rotatebox{270}{\includegraphics[width=8cm]{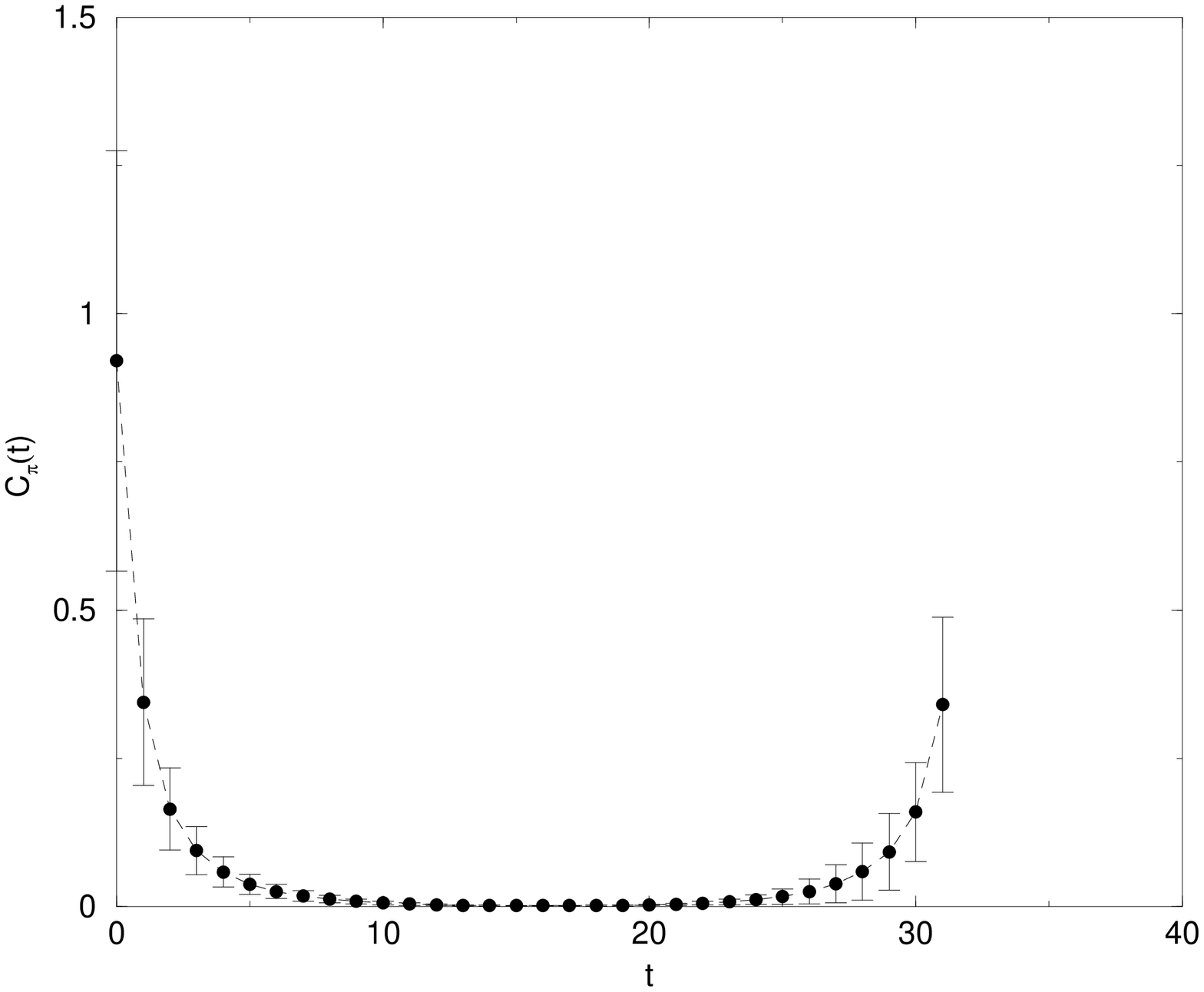}}
\end{center}
\caption{\label{pion_corr}  Green function of $\pi$
at $\beta=5.85$ and $\kappa=0.1585$ on the $8^3 \times 32$ lattice.  
The error bars represent statistical errors 
determined by the Jackknife method. }
\end{figure}

In Fig. \ref{pion_corr}, we 
show the pion 
correlation function at $\beta=5.85$, and $\kappa=0.1585$.
  In selecting the time range to be used in the fitting, we have tried
to be systematic. We choose the best fitting range by maximizing the confidence
level of the fit and optimizing $\chi^2/d.o.f$. 

\begin{figure}[hb]
\begin{center}
\rotatebox{270}{\includegraphics[width=8cm]{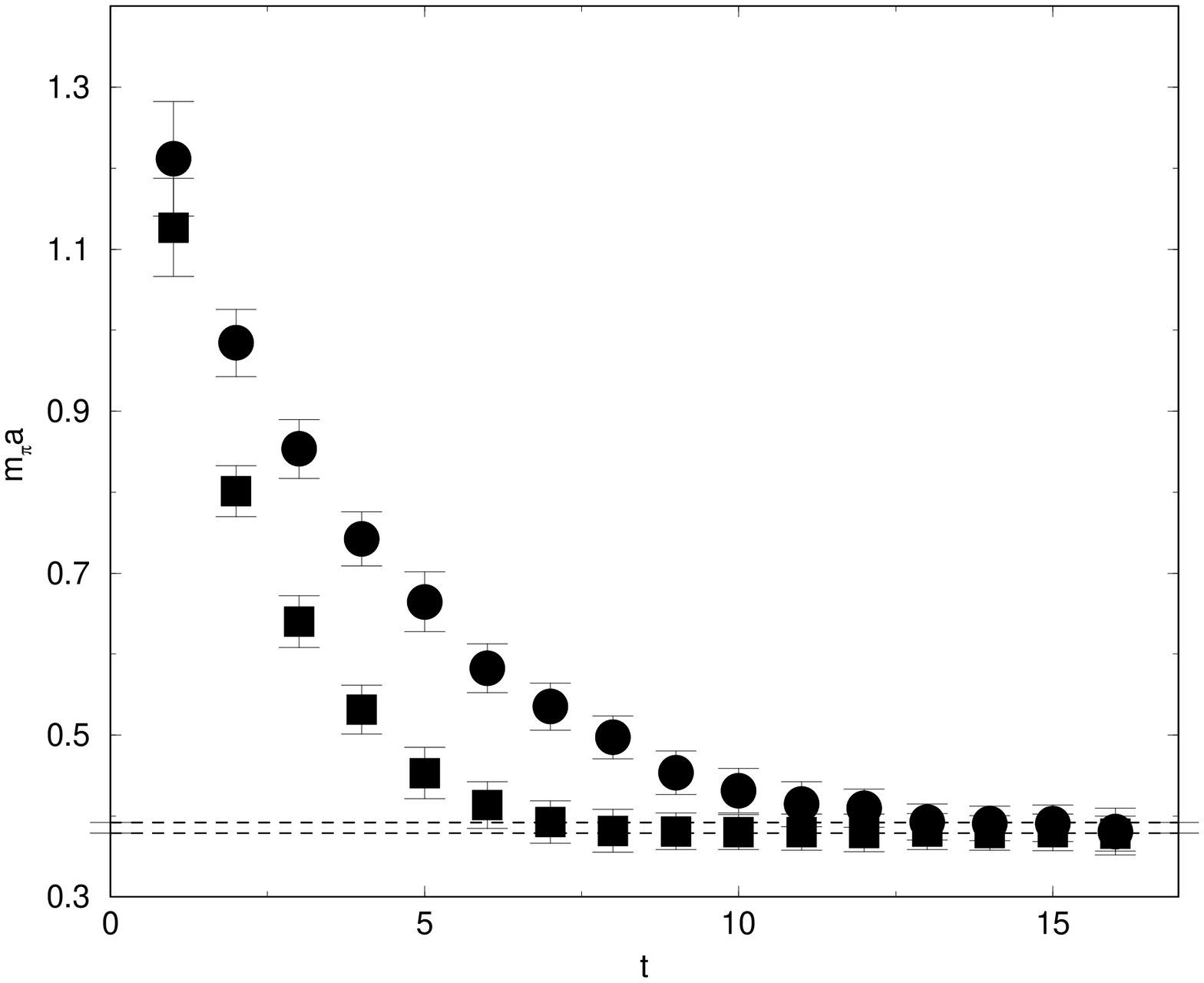}}
\end{center}
\caption{\label{pionsmea}  $\pi$ effective mass fits to 
the correlation function at  $\beta=5.85$ and $\kappa=0.1585$
and on the $8^3\times32$ lattice.
Data for the point source, smearing source for $r_0=1$, 
and $r_0=18$ are labeled 
by circles and squares respectively (from top to bottom).}
\end{figure}

\begin{figure}[hb]
\begin{center}
\rotatebox{270}{\includegraphics[width=8cm]{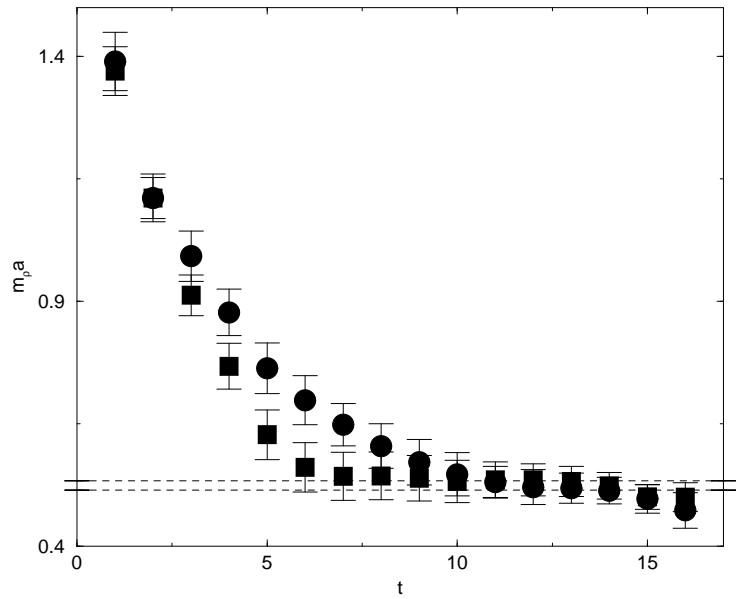}}
\end{center}
\caption{\label{rhosmear} The same as Fig. \ref{pionsmea}, 
but for the $\rho$ particle.}\end{figure}

\begin{figure}[hb]
\begin{center}
\rotatebox{270}{\includegraphics[width=8cm]{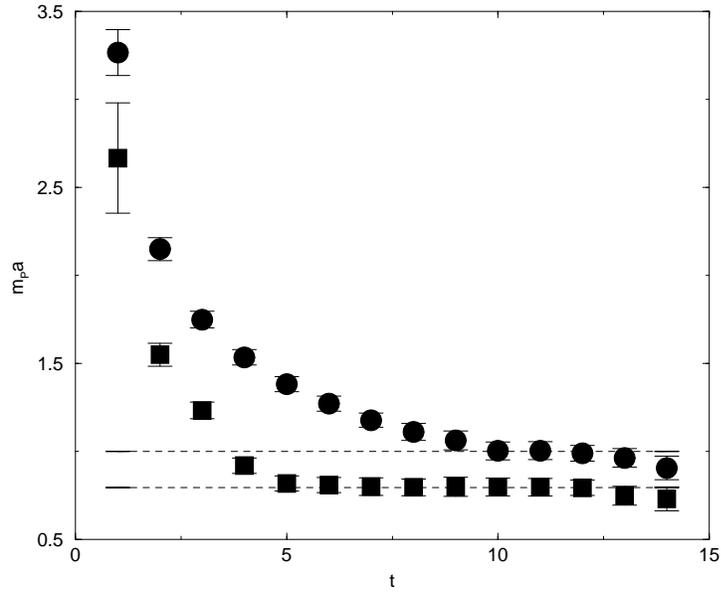}}
\end{center}
\caption{\label{protnsme} The same as Fig. \ref{pionsmea}, 
but for the proton particle.}\end{figure}

\begin{figure}[hb]
\begin{center}
\rotatebox{270}{\includegraphics[width=8cm]{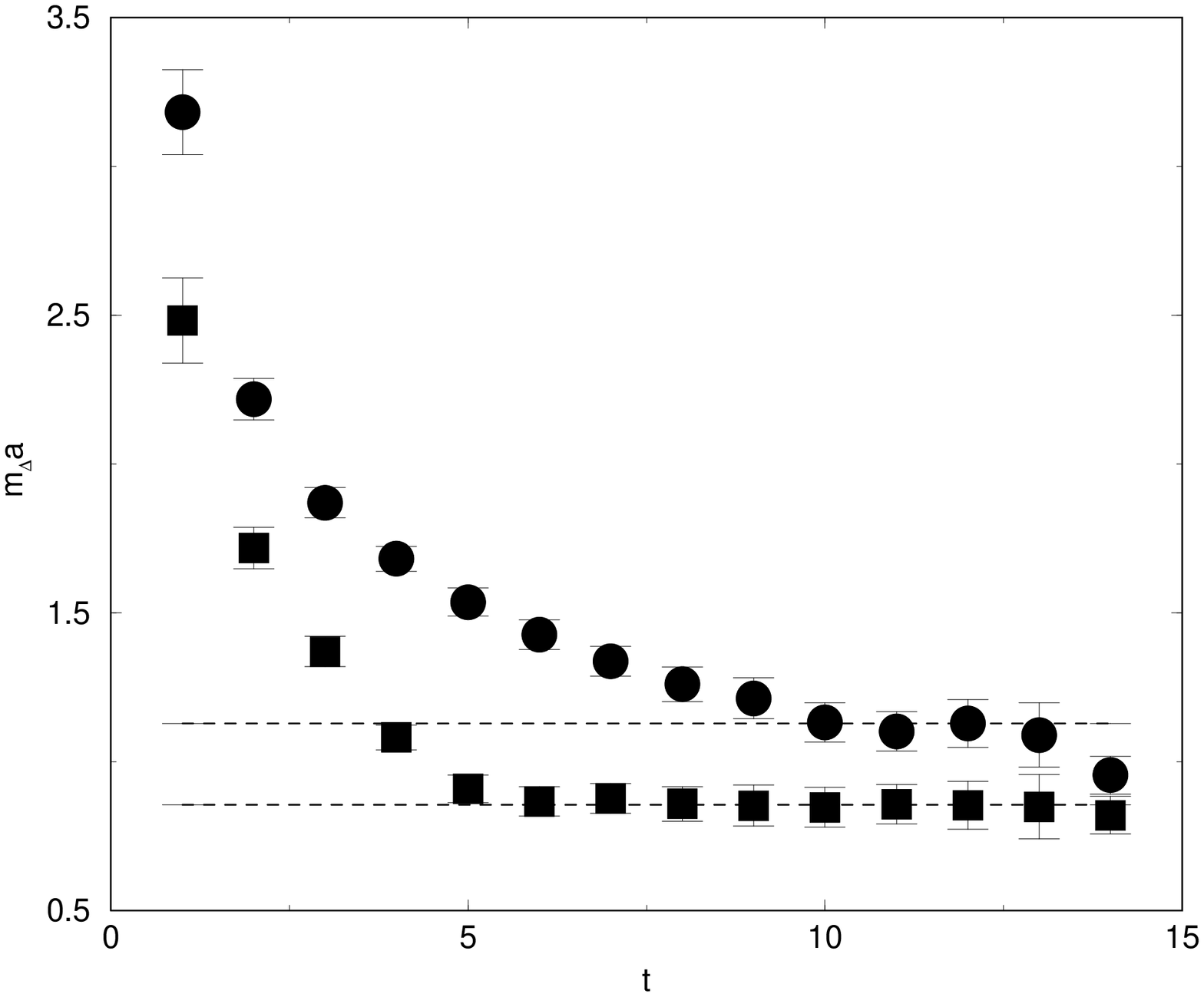}}
\end{center}
\caption{\label{deltasme} The same as Fig. \ref{pionsmea}, but for the 
$\Delta$ particle.}\end{figure}

Point source means a delta function, and smeared
source means a spread-out distribution 
(an approximation to the actual wave-function of the quantum state).
For example, the
simplest operator for a meson is 
just $O_h(\stackrel{\rightarrow}{x})={\bar q}(\vec{x}) q(\vec{x})$,
i.e. the product of quark and anti-quark
fields at a single point.
A disadvantage of this point source,  
is that this operator creates not only
the lightest meson, but all possible excited states
too.  To write down 
an operator which creates more of the single state, 
one must ``smear" the operator out, e.g.
\begin{eqnarray}
O_h(\vec{x})=\sum_{\vec{y}} ~ {\bar q} (\vec{x}) f(\vec{x}-\vec{y}) q(\vec{y}),
\end{eqnarray}
where $f(\vec{x})$ is some smooth function. Here we choose
\begin{eqnarray}
f(\vec{x}) = N\rm{exp}(-|\vec{x}|^2/r_0^2),
\end{eqnarray}
with $N$ a normalization factor.  
The size of the smeared operator
should generally be comparable to the size of the hadron created.
There is no automatic procedure 
for tuning the smearing parameter $r_0$.  One simply has to
experiment with a couple of choices. 
We plot  respectively 
in Figs. \ref{pionsmea}, \ref{rhosmear}, \ref{protnsme} and \ref{deltasme} ,  
the effective mass of $\pi$, $\rho$, proton and $\Delta$ particles,
as a function of time $t$ 
at $\beta=5.85$, and $\kappa=0.1585$ on the $8^3\times32$ lattice. 
As one sees, the plateau from which one can estimate the effective mass, 
is very narrow for point source, due to the reason mentioned above.
When the smearing source is used, 
the width of the plateau changes with the smearing parameter $r_0$.
We tried many values of $r_0$ and found that when $r_0\ge 16$, 
the effective mass is almost independent on $r_0$ 
where we observe the widest plateau.
These figures imply that the smearing technique plays more important role
for heavier hadrons
to suppress the contamination of the excited states;
Furthermore, one has to do careful study using the smearing technique, 
before doing simulation on a larger lattice.

We show the effective masses $am_h$ of the light hadrons 
in Tab. \ref{Tab3} with smearing parameter $r_0=18$. 
The best fits to a range of points begin at $t_{min}$=8 to $t_{max}$=16. 
The masses are good agreement with the MILC previous results on the
$16^3\times32$ lattice\cite{Bitar:1992dk}. 
This means that finite size effects are small
at this $\beta$ and $\kappa$.

\begin{table} 
  \begin{center}
  \begin{tabular}{|c|c|c|c|r|r|c|c|c|}\hline
   Particle  &  Group  & Configs & Lattice & $t_{min}$  & $t_{max}$  &   Mass   &  $\chi^2/dof$  &  $C.L.$  \\ \hline
   $\pi$ &   MILC   &  90 & $16^3 \times 32$ & 7   &       16 &     0.378(2) &     12.38/8    &  0.135  \\
        &   ZSU    & 200 &  $8^3 \times 32$ & 8   &       16 &     0.379(6) &     8.74/7    &  0.163   \\ \hline
   $\rho$ &    MILC   & 90 &  $16^3 \times 32$ & 8    &      16 &     0.530(3) &    2.857/7    &  0.898    \\
       &    ZSU    &  200 & $8^3 \times 32$ &  8     &     16 &     0.533(7) &      4.67/7    &  0.216   \\ \hline
   proton & MILC   & 90 & $16^3 \times 32$ &  7   &       16  &    0.783(10) &    8.339/8    &  0.401     \\
       &    ZSU    &  200 &  $8^3 \times 32$ & 8   &       16  &    0.796(19) &    11.36/7     &  0.112     \\ \hline
   $\Delta$ & MILC    & 90 & $16^3 \times 32$ &  8      &    16  &    0.852(11)  &   9.302/7    &  0.232      \\
        &   ZSU     & 200 & $8^3 \times 32$ & 8    &      16  &    0.857(13)  &  16.51/7     &  0.023 \\ \hline               \end{tabular}
          \end{center}
\caption{\label{Tab3} Effective masses of light hadrons at $\beta$=5.85 and $\kappa$=0.1585 on the 
lattice $16^3\times32$(MILC) and $8^3\times32$ (ZSU, this work).}
\end{table}

\begin{figure}[hb]
\begin{center}
\rotatebox{270}{\includegraphics[width=8cm]{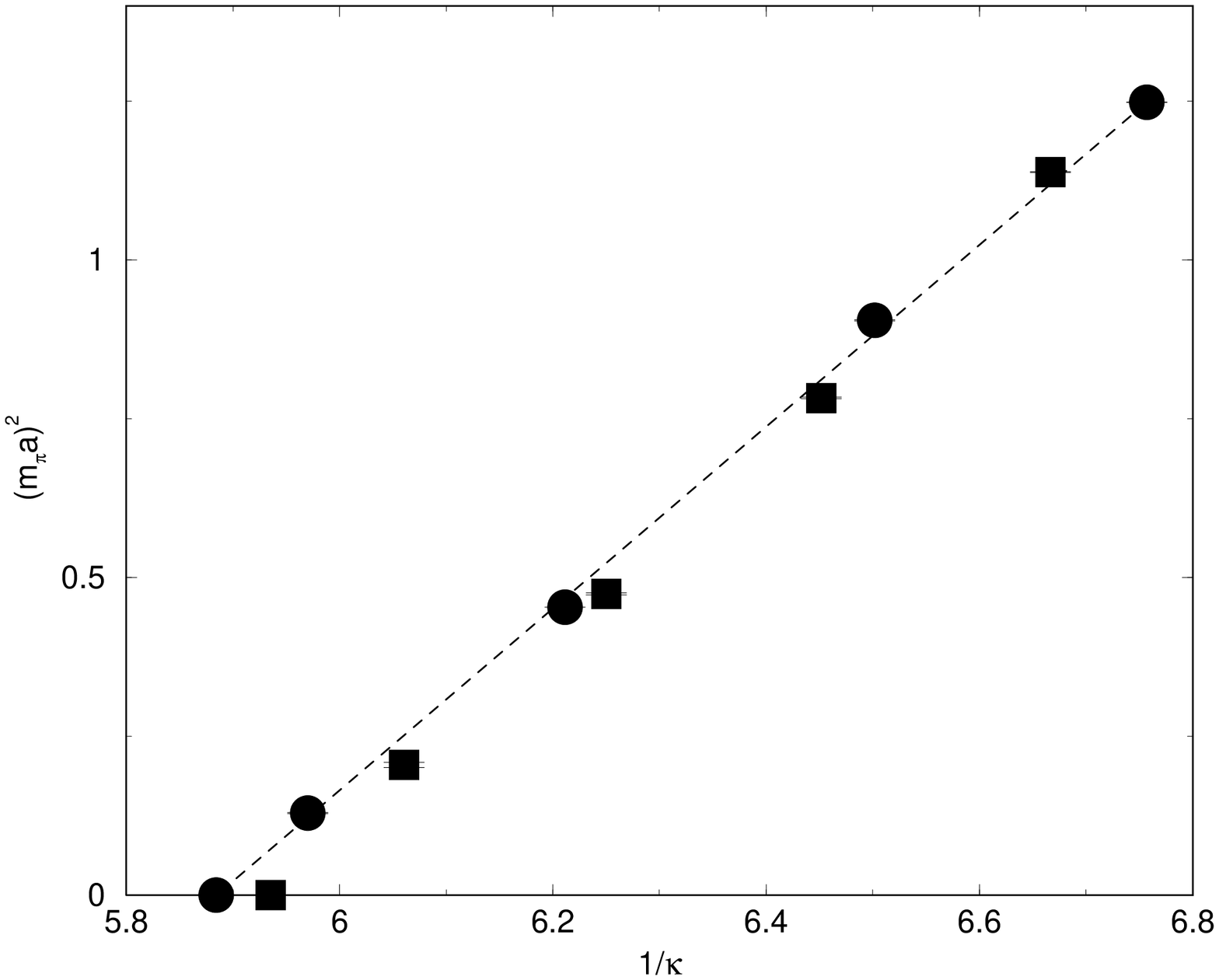}}
\end{center}
\caption{\label{mp2k_z_g} Pion mass squared as a 
function of $1/\kappa$ for $\beta$=5.7.
ZSU's and GF11's results are labeled by circles and squares.
  }
\end{figure}

\begin{figure}[hb]
\begin{center}
\rotatebox{270}{\includegraphics[width=8cm]{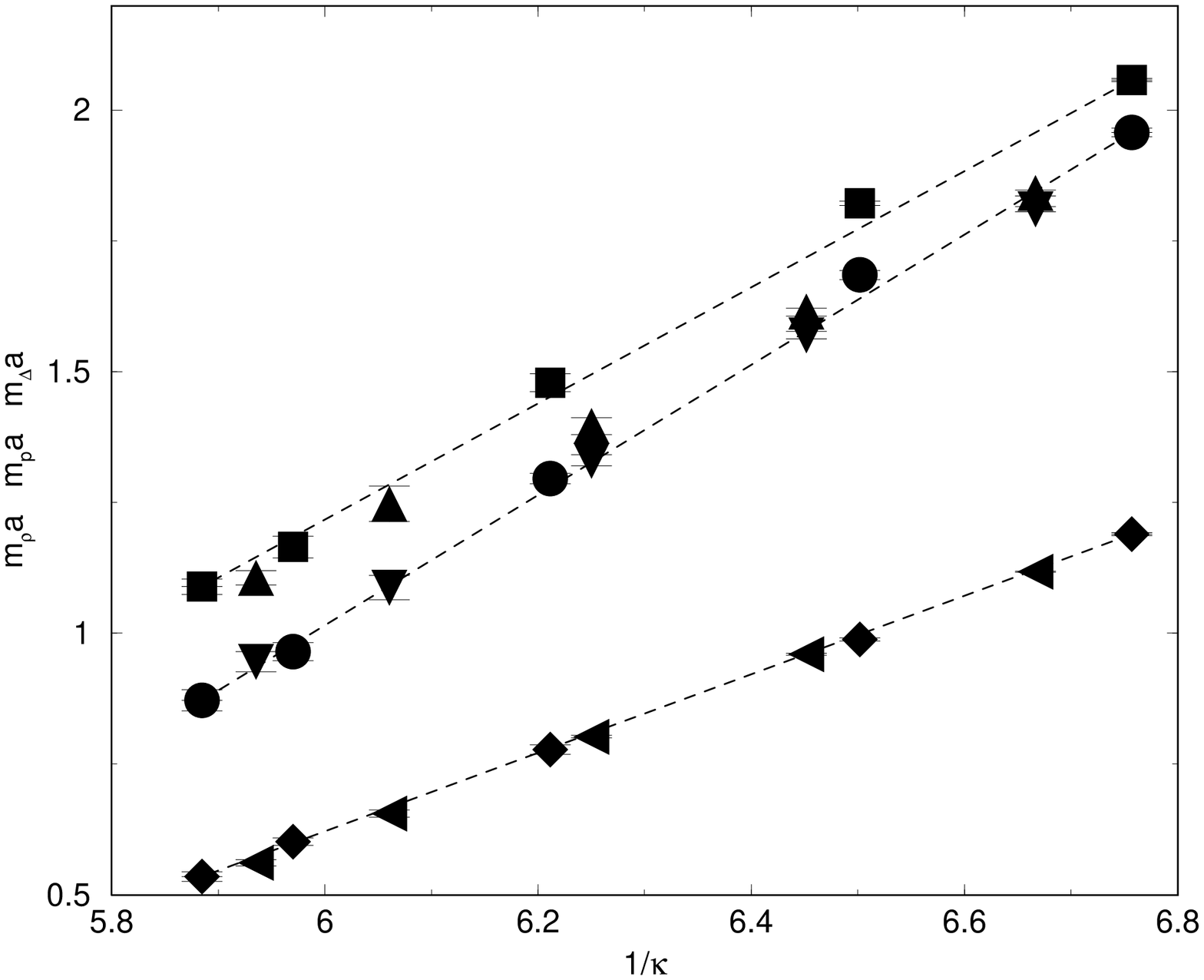}}
\end{center}
\caption{\label{mrpdk_z_g} Effective mass of $\rho$ 
(GF11: triangle left, ZSU: diamond),
proton (GF11: triangle down, ZSU: circle), $\Delta$ (GF11: triangle up,
ZSU: square) as a 
function of $1/\kappa$ for $\beta$=5.7. The points at the smallest value of $1/\kappa$ is the ZSU result extrapolated to the chiral limit.
  }
\end{figure}

In Figs. \ref{mp2k_z_g} and \ref{mrpdk_z_g},  
we compare our results (with $r_0=22$). 
for $\pi$ mass squared, $\rho$ mass, proton mass
and $\Delta$ mass  as a function
of $1/\kappa$ for $\beta=5.7$ with GF11
\cite{Butler:em} on the same $8^3\times32$ lattice.
The GF11 collaboration has 2439 configurations.
Most results are consistent.

To determine the relation between the lattice spacing $a$ and coupling
$\beta$, one has to input the experimental value of a hadron mass (see
Ref. \cite{Mei:2002fu} for details).

\subsection{$a_1 (P)$ and $1^{-+}$ hybrid meson masses}

At $\beta=5.85$ on the $8^3\times 32$ lattice, 
120 stored pure gauge configurations (see Tab. {parameters}) were re-used to study
$a_1 (P)$ and $1^{-+}$ hybrid meson masses. 
BiCGstab algorithm was employed to compute
the quark propagators with Wilson fermions and the residue is of $O(10^{-7})$.
Then we computed the correlation function using the sources and sinks 
in Tab. \ref{tab0}, from which the effective mass is extracted.
Our results at $\kappa=0.1450$ and $r_0=16$
are listed in Tab. \ref{Tab4}, and compared with
the MILC data\cite{Bernard:1997ib}.

\begin{table}
  \begin{center}
  \begin{tabular}{|c|c|c|c|c|c|c|c|}\hline
   Group & $\kappa$ &Configs & Lattice & Source(s)$\rightarrow$Sink         &  Fit Range & $ \chi^2/dof $ & Mass  \\ \hline
   MILC  &  0.1450 & 23 & $20^3\times48$ & $a_1(P)\rightarrow a_{1}(P)$      &     6-11     &       1.7/4   & 1.312(8)   \\
         &     & &    & $1^{-+}\rightarrow 1^{-+} $       &     4-10     &       3.5/5   & 1.88(8)     \\
         & & &        & $Q^{4}\rightarrow 1^{-+} $        &     3-7      &       0.7/3   & 1.65(5)      \\ \hline
   ZSU   &  0.1450 & 120 &$8^3\times32$ & $a_1(P)\rightarrow a_{1}(P)$      &     6-11     &       1.5/4   & 1.318(6)   \\
         &    & &     & $1^{-+}\rightarrow 1^{-+} $       &     4-10     &       4.2/6   & 1.87(10)     \\
         &       & &  & $Q^{4}\rightarrow 1^{-+} $        &     3-7      &       1.2/3   & 1.65(2)      \\ \hline        
                         \end{tabular}
          \end{center}
\caption{\label{Tab4} Effective masses for the ordinary P-wave $a_{1}$ meson 
and the exotic $1^{-+}$ meson for $\beta=5.85$ between MILC
and ZSU.} 
\end{table}

\section{Conclusions}

We have demonstrated that a parallel cluster of PC type computers is an 
economical way to build a powerful computing resource for academic purposes.
On an MPI QCD benchmark simulation it compares favorably with other MPI 
platforms. 

We also present results for the light hadrons and $1^{-+}$ hybrid meson 
from lattice QCD.
Such large scale simulations had usually required supercomputing resources,
but now they were all done on our PC cluster.
A more careful and systematic study of the smearing method is made.
Our results for $\beta=5.7$ and $5.85$ 
are consistent well with the data obtained on
supercomputers by other groups on the same or larger lattices.
This implies that finite size effects are small at these $\beta$ values.

To compare the lattice results with experiment, 
one needs to do simulations at larger $\beta$ and 
carefully study the lattice spacing errors. 
According to the literature, there are strong finite size effects 
for the Wilson action at $\beta \ge 6.0$ and  
very larger lattice volume is required. 
In this aspect, it is more efficient to use the 
improved action and some progress 
has been reported in Refs. \cite{Mei:2002ip, Luo:2002rz}.

In conclusion,  we are confident that ZSU's Pentium 
cluster can provide a very flexible and extremely economical computing 
solution, which fits the demands and budget of a developing lattice field 
theory group. 
We are going to use the machine to produce more
useful results of non-perturbative physics.


This work was in part based on the MILC collaboration's public 
lattice gauge theory code. 
(See reference \cite{milc}.)  We are grateful to 
C. DeTar, S. Gottlieb and D. Toussiant for helpful discussions.

\end{document}